\documentclass[
twocolumn,
]{ceurart}

\usepackage{listings}
\usepackage{makecell}
\usepackage{hyperref}
\usepackage{multirow}
\usepackage{array}
\begin{document}

\copyrightyear{2023}
\copyrightclause{Copyright for this paper by its authors. Use permitted under Creative Commons License Attribution 4.0 International (CC BY 4.0).}
\conference{Fifth Knowledge-aware and Conversational Recommender Systems (KaRS) Workshop @ RecSys 2023, September 18--22 2023, Singapore.}

\title{Tidying Up the Conversational Recommender Systems' Biases}


\author[1, 2]{Armin Moradi}[%
email=armin.moradi@mila.quebec,
]

\author[1, 2, 3]{Golnoosh Farnadi}[%
email=farnadig@mila.quebec,
]
\address[1]{Mila, Quebec AI Insitute}
\address[2]{Université de Montréal}
\address[3]{McGill University}


\begin{abstract}
The growing popularity of language models has sparked interest in conversational recommender systems (CRS) within both industry and research circles. However, concerns regarding biases in these systems have emerged. While individual components of CRS have been subject to bias studies, a literature gap remains in understanding specific biases unique to CRS and how these biases may be amplified or reduced when integrated into complex CRS models. In this paper, we provide a concise review of biases in CRS by surveying recent literature. We examine the presence of biases throughout the system's pipeline and consider the challenges that arise from combining multiple models. Our study investigates biases in classic recommender systems and their relevance to CRS. Moreover, we address specific biases in CRS, considering variations with and without natural language understanding capabilities, along with biases related to dialogue systems and language models. Through our findings, we highlight the necessity of adopting a holistic perspective when dealing with biases in complex CRS models. 

\end{abstract}

\maketitle

\begin{figure*}
\label{fig1:arch}
  \centering
  \includegraphics[width=0.7\linewidth]{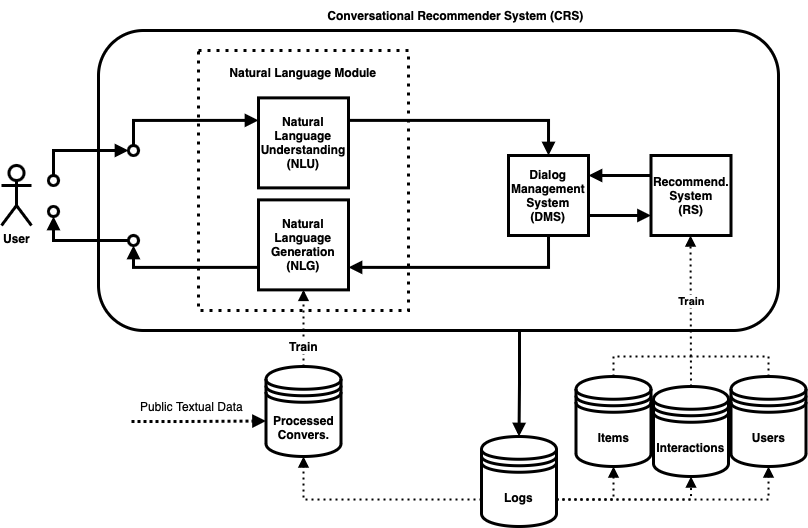}
  \caption{Architecture of the Conversational Recommender System. The diagram illustrates the intricate network of components, including a natural language module, dialogue management system, and recommender system working collaboratively to deliver personalized recommendations in a dynamic conversational interface.}
\end{figure*}

\section{Introduction}

In recent years, conversational recommender systems (CRSs) \cite{lei2020interactive, deng2022user, li2021self, li2022user, ren2022variational, radensky2023think, li2020ranking, zhou2020improving, jannach2022conversational} have garnered significant attention, reshaping personalized recommendations through interactive user engagements. This transformation is notably supported by the successful integration of large language models (LLMs) like ChatGPT \cite{ray2023chatgpt, dai2023uncovering}, thereby driving the widespread deployment of LLMs in various applications. Such models have found substantial integration in prominent platforms, including Microsoft Bing\footnote{\url{https://blogs.bing.com/search/march_2023/Confirmed-the-new-Bing-runs-on-OpenAI’s-GPT-4}}, which has invigorated dialogue search engines and recommender systems with unprecedented capabilities and paving the way for a new era of user engagement.

Although biases within recommender systems have garnered significant attention \cite{jin2023survey, milano2020recommender, gupta2021correcting, liu2022rating, lin2021mitigating, zhu2021popularity, melchiorre2020personality}, the examination of biases in conversational recommender systems remains a relatively unexplored domain \cite{Jannach_2021}. Despite the conceptual alignment between conventional recommender systems and their conversational counterparts, the latter exhibit heightened complexity and an increased potential for biases \cite{lin2022towards}. It is worth noting that some existing research has specifically addressed biases in conversational recommender systems \cite{ li2021self, li2022user, ren2022variational, radensky2023think, lin2021mitigating, lin2022towards, shen2022unintended}. However, no concerted effort has been undertaken to systematically categorize and analyze biases unique to conversational recommender systems, including how previously studied biases manifest within the conversational context. This paper aims to bridge this gap by offering a preliminary exploration into the intricate biases that characterize these complex systems.

This study conducts a systematic literature review to explore biases within conversational recommender systems. The methodology involves analyzing recent papers from top conferences in machine learning and information retrieval. Keyword searches within titles and abstracts identify relevant contributions that shed light on biases, including fairness concerns and bias amplification. 

The paper begins by thoroughly investigating biases in classic recommender systems, establishing a foundation for understanding common biases, addressing similar notions to each of them, and most importantly, examining each bias in a conversational setting in Section~\ref{sec1}. This step ensures a holistic grasp of biases across traditional recommendation systems. We then delve into each bias within CRSs. We start with focusing on CRSs without natural language understanding which uses basic dialogue systems for user interaction in Section~\ref{sec2}. Furthermore, to capture diverse aspects and potential biases arising from natural language understanding, a dedicated literature review is conducted on the more complex CRSs that aim to understand natural language in Section~\ref{sec3}. Following that, in Section~\ref{sec:future}, we present the limitations and potential avenues for future research of our study. Finally, in Section~\ref{sec:conclusion}, we draw conclusions and wrap up the paper.

\section{Related Work}

Although biases in general machine learning models, natural language processing (NLP), and recommender systems (RS) domains have undergone extensive study, biases in conversational recommender systems (CRSs) have received limited attention.

On a broad aspect, ML biases have been a subject of interest, with an example of the work by Mehrabi et al. \cite{mehrabi2022survey} which provides a comprehensive survey on the challenges and approaches to mitigate biases in ML models. Additionally, NLP biases have received attention, and for instance, Blodgett et al. \cite{blodgett2020language} discussed the manifestation and implications of biases in language models which can be seen as an important part of CRSs.

Moreover, several surveys have focused on biases and fairness in recommender systems (RS) as a whole \cite{jin2023survey, wang2023survey, chen2021bias, milano2020recommender}. These surveys provide valuable insights into biases present in RS, which is a sub-module of CRSs, therefore they can be used as valuable sources for specific investigation in a conversational setting.

Within the realm of CRS biases, recent research has shed light on different biases that can arise in such systems. For example, unintended biases are discussed by Shen et al. \cite{shen2022unintended}. Lin et al. \cite{lin2022quantifying} quantified biases in CRS by exploring the fairness of recommendations across different user groups. Another study by Lin et al. \cite{lin2022towards} highlighted the importance of addressing biases in CRS to ensure equitable and inclusive recommendations.

As for CRS surveys, Jannach et al. \cite{Jannach_2021, Jannach_2022, jannach2022conversational} have contributed significantly to the understanding of conversational recommender systems. However, to the best of our knowledge, a dedicated survey paper focusing solely on the biases in conversational recommender systems is still missing from the literature. Consequently, this paper aims to embark on the initial journey towards gaining a deeper understanding of the biases present in CRS and their interactions, elucidating how biases from various components within this complex system can either be accentuated or alleviated.

\section{Methodology}

The primary objective of this study is to conduct a review of existing literature on potential biases in conversational recommender systems. To achieve this goal, a systematic approach is adopted, focusing on papers published in prominent machine learning and information retrieval-related conferences from January 2019 to June 2023. These conferences include knowledge discovery and data mining (KDD), Special Interest Group on Information Retrieval (SIGIR), ACM Conference on Recommender Systems,  User Modelling, Adaptation and Personalization (UMAP), International World Wide Web Conference (WWW), Neural Information Processing Systems (NeurIPS), International Conference on Machine Learning (ICML).

To identify relevant papers, we conducted keyword searches within the titles or abstracts of the papers presented at these conferences. The selected keywords include bias, dialog, conversation, chat, question, mitigat*, recommend*, amplif*, fair*. These were chosen to align with key aspects within the scope of biases in conversational recommender systems and related notions such as dialogue systems and classic recommender systems. These keywords were thoughtfully chosen to reflect the themes within the scope of biases in conversational recommender systems. Specifically, they align with key aspects such as recommender systems, dialogue systems, fairness considerations, and the amplification of biases. The chosen keywords collectively encompass a wide spectrum of research that pertains to these facets, ensuring that our selection is representative of the relevant literature landscape. Subsequently, we filtered the results once again based on their contributions to the trustworthy intricacies of recommender systems, conversational recommender systems, or dialogue systems. 

\section{Conversational Recommender Systems}


To explore biases within conversational recommender systems (CRSs), a precise definition and understanding of their model architecture are crucial. A Conversational Recommender System (CRS) is intricate, with interconnected components (Figure \ref{fig1:arch}). Users interact with the system to either provide answers (feedback) to the system's questions (recommendations) or receive personalized recommendations. Within the CRS, a language module assumes a pivotal role, consisting of two submodules: natural language understanding (NLU) and natural language generation (NLG). The NLU empowers the system to understand user intentions, extracting insights from their input and prior profiling data. On the other hand, the NLG crafts coherent and contextually relevant responses in natural language. Alongside this, a recommender system undertakes user queries, harnessing available data to generate personalized recommendations. 

The dialogue management system (DMS) as the heart of the model, is in charge of orchestrating conversations between the user and the system, ensuring logical flow and pertinence in each interaction. At each point of the dialogue, The DMS decides when to finalize recommendations or seek more information, guided by NLU and the recommender system and by navigating the outputs through the NLG. Continual system enhancement is achieved through the accumulation of conversation logs into a database, leveraging these processed logs to train both the natural language module and the recommender system. This iterative training process augments their capabilities over time, refining user interactions. 

It's important to emphasize that the architecture outlined above represents the general structure of a CRS. However, certain CRSs do not incorporate natural language understanding and generation into their operations; instead, they adopt simplified input processing methods. This leads to the categorization of CRSs into two types: \emph{Topic-guided CRSs}, which utilize natural language understanding, and \emph{Attribute-aware CRSs}, which rely on simplified input processing methods and attribute inquiries, as classified by Ren et al. \cite{ren2022variational}.

Another significant point to consider is that a traditional recommender system can also be illustrated by utilizing a subset of the modules featured in Figure \ref{fig1:arch}. By removing the Natural Language Module and the Dialogue Management System (DMS) components, and by establishing a direct communication path connecting the Recommender System and the User, we can achieve a simplified structure. This approach facilitates a clearer understanding of the underlying nature and impact of biases discussed in Section \ref{sec1}.

\section{Biases in Classic Recommender Systems}
\label{sec1}

In this section, we examine biases in classic recommender systems across three key aspects for each bias. For each bias, first, we define and discuss the bias, drawing from relevant literature for a strong foundation. Second, under \textit{Similar Notions}, we identify related notions that fall within each bias. And third, in \textit{Through the CRS lens}, we analyze each bias in the context of conversational recommender systems, exploring specific works in this domain. This approach offers a wide-ranging perspective on biases in classic recommender systems within conversational interactions.


\subsection{Popularity Bias}
Popularity bias in recommender systems prioritizes popular items, assuming they are more likely to interest users. However, this bias can lead to a lack of diversity, overshadowing lesser-known options.  Recognizing and mitigating popularity bias is crucial for developing recommender systems that offer a wider range of choices and promote serendipitous discovery \cite{chen2021bias, brown2022diversified}. Besides investigating and mitigating this bias in a classic fashion, there are some other works that focus on different settings and notions. For example, Zhu et al. \cite{zhu2021popularity, zhang2023invariant} try to solve the popularity bias in a dynamic environment setting. Abdollahpouri et al. \cite{abdollahpouri2020connection} see how different types of users are affected by popularity bias and in the study by Zhu et al. \cite{zhu2021popularity} the focus pertains to the challenge of ranking a set of equally favored items based on their popularity.\\
\textit{Similar notions:}
\begin{itemize}
    \item  \textbf{Long-tail bias:} The opposite of popularity bias can also be the issue in a model, over-recommending the niche items instead of already established items \cite{abdollahpouri2019managing}.
    \item  \textbf{Filter bubbles, Echo chambers, Polarization:} These concepts constitute a significant area of research within recommender systems and often arise as consequences of existing popularity bias. In the work by Michiels et al. \cite{michiels2022filter}, the filter bubble is characterized by a ``decrease in any dimension of diversity." It's essential to recognize that this notion doesn't solely originate from popularity bias; various other biases can contribute as well. Wang et al. \cite{wang2022user} present a user-controllable model to alleviate filter bubbles, which holds potential for application in conversational settings.
\end{itemize}
\textit{Through the CRS lens:} Lin et al. \cite{lin2022towards} examine the prevalence of popularity bias by investigating three CRS models as baselines, revealing its existence in these systems as well. It is speculated that this bias can be alleviated through the conversational setting because the user is capable of criticizing the recommendations in order to have a more niche recommended list of items. 


\subsection{User Log Bias}
Different underlying biases can lead to a biased log of user data. Therefore we define User Log bias as an umbrella term for the discrepancy between real-world representation of user-item interactions (the ground truth) and the recorded data. This variation can happen due to many reasons, such as oversimplifying the user logs, not having access to certain user data and imprecise user-item interaction modeling. 

There are some works that address this bias with different approaches and different but similar bias definitions. Frumermann et al. \cite{frumerman2019all} investigate the real meaning behind rejected items and whether all the rejected items should be seen as the same. On the same issue, Nazari et al. \cite{nazari2022choice} make an effort to use user-implicit signals and Xu et al. \cite{xu2022ukd} try to leverage unclicked items in the dataset in addition to the interactions. Lastly, Zhang et al. \cite{zhang2022counteracting} focus on how we should tackle user inattentiveness to the items that are being interacted with but do not necessarily correlate with user satisfaction.\\
\textit{Similar Notions:}
\begin{itemize}
    \item \textbf{Conformity Bias:} It is a cognitive bias which is defined as users' disinclination towards negatively rating an item because of the item's high ranking or popularity \cite{chen2021bias, zheng2021disentangling}.
    \item \textbf{Exposure Bias:} Since each user is randomly exposed to a subset of items during her lifetime, her profiling is biased towards the items that they have already been exposed to \cite{gupta2021correcting, dash2021umpire, mcinerney2020counterfactual}. Also, it indicates that unobserved items do not always represent negative preference \cite{chen2021bias}.
    \item \textbf{Selection Bias:} It is a similar notion to exposure bias, as it is defined as the un-randomness of the missing interactions \cite{zhang2020large, ovaisi2020correcting, liu2022rating}.
    \item \textbf{Exploitation Bias:} It is investigated that users have a tendency to interact with or rate items that they personally prefer \cite{huang2020keeping, yang2022can}.

\end{itemize}
\textit{Through the CRS lens:} 
Likewise, biases in conversational settings concerning user logs have been examined, particularly regarding the process of dataset curation. Szpektor et al. and Yu et al. \cite{szpektor2020dynamic, yu2019visual} note that the limited number of individuals responsible for labeling conversation quality may introduce bias stemming from their personal preferences. Additionally, Pang et al. \cite{pang2023auditing} highlight their focus on cultural differences, which can result in divergent labelings of CRS datasets.
 
Looking back to the \textit{similar notions}, conformity bias is also studied in the conversational recommender systems \cite{raul2023cam2, yang2023debiased}. Overall, the conversational nature of CRSs and the inevitable increasing complexity can cause the model to have more biases in the log than the classic models leading to a reduction of the quality of the datasets which needs to be addressed in future studies.

\subsection{Recommendation Evaluation Bias}

In order to evaluate a recommender system, certain priorities need to be addressed with respect to the needed specifications of the system in addition to common accuracy and ranking evaluation metrics. For example, serendipity and diversity of the recommendations and long-term vs short-term fairness can all be taken into account as a way to measure the recommender system \cite{d2020fairness, wen2022distributionally, mladenov2020optimizing}. Therefore, it is important to establish metrics to assess the various aspects or qualities of a recommender system that need to be evaluated. Also, the flaws of some of the already established metrics have been challenged \cite{hiranandani2020optimization, christakopoulou2020deconfounding} and some even propose a metric-less offline evaluation method \cite{diaz2022offline}. Additional metrics such as evaluating fairness in ranking \cite{raj2022measuring}, and recommendation uncertainty \cite{cohen2021not} are also addressed in the literature. Building on fairness, De et al. \cite{de2023unfair} go beyond fairness metrics and suggest that search engines can manipulate users while maintaining top-notch fairness metrics.  Lastly, Wang et al. and Zhang et al. \cite{wang2019social, zhang2022counteracting} try to investigate user attention and how it affects recommendation models and the way users interact with it, directly influencing the ways we measure their capability.\\
\textit{Through the CRS lens:} There can be various use-cases for conversational recommender systems and different measurements can and should be prioritized depending on them. This makes choosing the evaluation metrics for conversational recommender systems a challenging task. For example, Lin et al. \cite{lin2022towards} propose that the Success Rate metric does not indicate how much the recommender system is benefiting each of its individual users. When evaluating a topic-guided conversational recommender system, specific metrics could come into play, such as psychologically inspired measures and those assessing conversation quality and engagement \cite{ghandeharioun2019approximating}. These metrics also address sub-objectives like accurate user satisfaction estimation \cite{deng2022user, sun2021simulating}. Lastly, it is very important to evaluate the natural-language-understanding CRSs in order to measure the gap between the understanding of the NLU and how much this understanding is utilized by the recommender systems \cite{wang2022towards}. One of the fundamental reasons is explored in the study by Zho et al. \cite{zhou2020improving}, where they investigate the disparity between the natural language representation of a potentially recommended item and its lack of precise alignment.

In the realm of CRS evaluation, the incorporation of new aspects amplifies complexity. Evaluating different modules and the complete CRS requires meticulous consideration, making CRSs more susceptible to recommendation evaluation biases.


\subsection{Attribute Bias}

Attribute bias is a concerning issue in recommender systems. These systems can inadvertently amplify existing societal biases by making recommendations based on certain sensitive attributes that each user can have, such as gender, race, or age. This bias can lead to unfair and discriminatory outcomes, as individuals from certain demographics may be systematically excluded or receive less favourable recommendations. Addressing demographic bias in recommender systems is essential for ensuring equal and equitable treatment for all users, regardless of their demographic characteristics.\\
\textit{Similar notions:}
\begin{itemize}
\item \textbf{Demographic bias} and \textbf{minority bias} can also be utilized to refer to Attribute bias \cite{islam2019mitigating, ying2023camus}.
\item \textbf{User Activity Bias:} There are several papers discussing the disparate impact of active users on the model \cite{eskandanian2019power}, and how differently the model interacts with them \cite{li2021user}.
\item \textbf{Sentiment Bias:} It is investigated that the more the users have positive interactions with the system they are more likely to get higher quality recommendations \cite{lin2021mitigating}.
\end{itemize}
\textit{Through the CRS lens:} Due to the potential usage of natural language processing in conversational models, other attributes of users can be exposed to the model and can be exploited. For instance, in a voice dialogue system, the accent of the user is a sensitive attribute that should ideally not influence the system's decision-making \cite{accentgap2018}. Also in the work by Cogswell et al. \cite{cogswell2020dialog}, it is discussed that the modality of presenting the data can affect the minorities' ability to perceive it. On a similar issue, different people interact differently with regard to their age and how experienced they are in interacting with a recommender system \cite{zheng2022ddr} and the model should be robust in different interactions and be able to extract the users' needs without emphasizing their demographic attributes.

In a conversational setting, user interaction empowers them to navigate and refine recommendations, allowing them to mitigate attribute biases in the final list, similar to how they can address popularity bias. Nevertheless, additional research is essential to examine which users' attribute biases can be effectively mitigated through conversation, and also to understand which biases, such as gender or race bias, might be exacerbated due to the inherent biases of various components within CRS.


\subsection{Position Bias} 
It happens as users tend to interact with items in higher positions of the recommendation list regardless of the items’ actual relevance so that the interacted items might not be highly relevant \cite{chen2021bias, ruffini2022modeling}. 

\textit{Similar Notions:}
\begin{itemize}
\item It is in accordance with \textbf{lead bias} in a work by Zhu et al. \cite{zhu2021leveraging} for news recommender systems that show how the `lead' part of the news in the recommender systems can bias a user's behavior towards the item.
\end{itemize}
\textit{Through the CRS lens:} In a conversational setting, there are options to counter position bias, such as providing explainability for each recommended item and framing the recommendations using natural language. These strategies can potentially mitigate the impact of position bias. It's important to note that this bias is intertwined with Framing bias (Section \ref{sec3}), as the ranking can be seen as a framing of data presentation and position bias can be considered a form of framing bias as well. Nevertheless, it is crucial to acknowledge that explanation methods which rely on an additional or surrogate model to provide justifications for why specific items are recommended to the user and ranked higher, are also susceptible to biases. Consequently, these explanations might not accurately reflect the performance of the original model.


\subsection{Personalization Bias}

Personalization bias in recommender systems refers to the tendency of these systems to continuously recommend similar content based on a user's preferences, potentially limiting their exposure to diverse perspectives and new experiences. Balancing personalization with serendipity is crucial to mitigate this bias and ensure users are presented with a broader range of recommendations \cite{chen2021bias}.\\
\textit{Similar Notions:}
\begin{itemize}
    \item \textbf{User preference Amplification:} Kalimeris et al. \cite{kalimeris2021preference} talk about how even relevant and high-quality recommendations can lead to user preference amplification, therefore, decreasing the users' exposure to diverse content.
    \item \textbf{Feedback Loop:}  User tends to follow the recommendations and the recommendations become the user’s interests themselves, leading to amplified biases. Moreover, in the long run, and with repetition of this loop, the amplified biases become the ground truth as the user logs are utilized to train other models \cite{mansoury2020feedback}.
\end{itemize}
\textit{Through the CRS lens:} Similar to the Popularity bias, in a conversational setting, users' ability to navigate the recommendations after receiving them gives them the option to reduce the attribute biases in the final list of recommendations as well.

With the inclusion of additional data such as conversations in user interactions, the model's tendency to overemphasize the previous dialogues may increase, potentially exacerbating personalization bias. Therefore, the presence of diverse data sources should be carefully managed to strike a balance between personalization and diversity in recommendations.


\subsection{Context Bias}

In a work by Zheng et al. \cite{zheng2022cbr}, the concept of "context bias" is explored as a comprehensive framework for analyzing a collection of recommended items. The study highlights that users' decision-making processes can be influenced by a combination of biases when presented with options that possess different attributes and potentially unique biases for each item. For instance, when browsing a news website, users may encounter various forms of content, such as text and video (modality bias), alongside popularity-driven recommendations (popularity bias). Understanding how this set of biases, collectively referred to as context bias, impacts decision-making requires a broader and more in-depth investigation.\\
\textit{Through the CRS lens:} In conversational settings, context bias can play a vital role as the conversational nature of the system makes it more complicated and the dynamics of the model and different existing biases should also be addressed in the same context. 
In a conversation, the concept of context assumes a broader and more intricate definition compared to its application in traditional recommender systems confined to lists of items. Therefore, the potential exacerbation of context bias becomes a relevant consideration when applied to CRSs.


\section{Biases of Attribute-aware Conversational Recommender Systems}
\label{sec2}


Conversational recommender systems introduce a new set of biases that can impact recommendations and user experiences. While traditional biases in recommender systems have been extensively studied, the conversational nature of these systems introduces new biases in distinct ways. In this section, we conduct investigations on the existing biases of conversational recommender systems that do not have natural language understanding and interact with the user based on inquiring about attributes that they want to make decisions upon, such as \cite{wang2022towards, lei2020interactive, li2021seamlessly, ren2021learning, xu2021adapting}.



\subsection{Anchoring Bias}
In a conversational setting, the model has the option to utilize the user information throughout the conversation, even through different sessions depending on the system. Therefore, it is a challenge to how this information and the user's behavior and feedback to different recommendations should be utilized. Anchoring bias, or User History bias, happens when the previous recommendations and the user dialogue history in the conversation can anchor subsequent recommendations, leading the system to focus on a particular subset of items and potentially overlooking other options, therefore it is vital to catch the dynamics of user profile while being able to make use of the information. There have been some studies related to this bias. It is investigated that the dynamic nature of conversational systems can amplify the impact of anchoring bias \cite{qu2019bert, li2022user}. Also, Ren et al. \cite{ren2022variational} investigate the differences between old and new user preferences and the way they change.


\subsection{Attribute Selection Bias}
Attribute selection bias, which can also be called User Preference Assumption bias, refers to a phenomenon in conversational recommender systems where the system becomes biased towards the attributes that users prioritize when making decisions \cite{lin2022towards}. In each step of the recommendation process, the system may assume that the user wants to base their choices on specific attributes, thereby influencing the recommendations accordingly. This bias can impact the diversity and fairness of the recommendations by potentially overlooking alternative attributes that users might value but are not explicitly expressed. By primarily focusing on a subset of attributes, the system may limit the scope of recommendations, potentially hindering serendipitous discoveries and failing to provide a comprehensive and personalized experience.


\subsection{Human-AI Interaction Bias}

AI-conversation bias refers to a type of bias where users alter their conversational behavior and speech patterns when interacting with a conversational recommender system that they know is powered by an AI bot \cite{folstad2021future}. When users are aware that they are conversing with an artificial intelligence rather than a human, they may consciously or unconsciously modify their language, tone, or style of communication. This bias can arise from various factors, including a perceived need to simplify language, adapt to the system's limitations, or conform to social norms associated with human-AI interactions. As a result, the quality and naturalness of the conversation may be affected, potentially leading to a less engaging and authentic user experience \cite{hosseini2020simple}.


\subsection{Modality Bias}
Modality bias in a conversational recommender system setting refers to the tendency of multi-modal text generation models, such as the multi-modal GPT-4 model \cite{openai2023gpt4}, to heavily rely on textual input while paying less attention to non-textual signals, such as visual cues or signals \cite{tian2023multi}. This bias can limit the system's ability to effectively incorporate and leverage non-textual signals, leading to a potential loss of valuable information and a less holistic understanding of user preferences. Addressing modality bias involves developing more balanced and comprehensive models that can effectively capture and utilize both textual and non-textual cues.




\section{Biases of Topic-guided Conversational Recommender Systems}
\label{sec3}

In this section, our focus is to investigate the biases of natural language understanding models and their potential implications when integrated into conversational recommender systems with natural language modules. These types of models can be considered as a general type of the existing CRSs in the literature \cite{zhou2020improving, sarkar2020suggest}.


\subsection{Defective Queries Bias}

Defective query bias in a conversational recommender system setting occurs when users intentionally manipulate the system or unknowingly express ambiguous statements, making it difficult for the system to comprehend the conversation and generate useful recommendations \cite{li2021self, lu2021streamsketch, shen2022distributional, wang2023zero, liao2022ptau, mirzaei2022question}. This poses a challenge as the system struggles to fully grasp user intent and preferences. In such situations, the system may need to ask clarifying questions to obtain additional context and disambiguate user queries \cite{montazeralghaem2021large, aliannejadi2019asking}. However, if the system fails to address this bias effectively, it may lead to the recommendation of low-quality items that do not align with the user's preferences. Consequently, users may need to provide feedback or criticize the initial recommendations to prompt the system to refine its list of suggested items \cite{yu2019visual, li2020ranking, yang2022toward}. Addressing Defective Query bias requires the development of robust and adaptive conversational recommender systems that can handle uncertainties, disambiguate user queries, and incorporate user feedback to improve the quality and relevance of recommendations.



\subsection{Cognitive Biases}
 Cognitive bias in language models within a conversational recommender system setting involves the evaluation of large language models in relation to the cognitive biases observed in humans. These biases have the potential to directly impact the system's natural language generation module, influencing it to make irrational decisions depending on whether or not it is affected by these cognitive biases. Detecting and understanding the presence of cognitive biases in language models is crucial to ensure that the recommendations provided are fair and unbiased. By addressing and mitigating these biases, conversational recommender systems can strive to deliver more objective and rational recommendations that are not influenced by human cognitive biases \cite{jones2022capturing}.\\
 \textit{Similar Notions:}
\begin{itemize}
    \item \textbf{Framing bias:}  Framing in conversational recommender systems refers to how information presentation influences user perception and decision-making. By addressing framing biases through transparent and balanced recommendations, systems can enhance fairness and effectiveness \cite{reiter2023exploration, mulder2021operationalizing}.
    \item \textbf{Uncertainty-aversion bias:} This bias arises from users' negative inclination toward uncertain recommendations \cite{park2021experimental}. It can impact the effects of explainability techniques on user behavior. Moreover, it can also be studied on the effects of uncertainty in manually labeling datasets.

    \item \textbf{User Trust Bias:} {User Trust Bias} It is been studied that having a conversational interface will increase the trust rate of users \cite{gupta2022trust, radensky2023think}. Karduni et al. \cite{karduni2021images} argue that the way the faces are shown in news posts affects how much the users trust the platform. Also, the proactivity of the bot is discussed in the works by Kraus et al. \cite{kraus2022including, kraus2020effects}, Zhu et al. \cite{zhu2021proactive} and Lei et al. \cite{lei2022interacting} and it is verified that it affects human trust.
\end{itemize}


\subsection{Unintended Bias}
 
Unintended bias in a conversational recommender system setting refers to the unintentional influence that certain factors or characteristics may have on the recommendations provided \cite{shen2022unintended}. These biases can arise from societal stereotypes, historical data biases, or implicit associations present in language modeling and recommendation algorithms. Unintended biases can result in unequal treatment or favoritism towards certain groups or preferences, leading to potentially unfair or biased recommendations.


\subsection{Persona Bias}

There is Persona bias in a dialogue system, and therefore in a topic-guided CRS, refers to the detrimental discrepancies in responses that arise when different personas are adopted. There are a few works in dialogue models and recommendation systems that try to adapt the generation process by conforming to the user's persona \cite{han2023persona, xu2022cosplay, gu2021partner, ma2021one, tigunova2019listening, fatahi2023investigating}.  These biases manifest in various ways, such as variations in the level of offensiveness or agreement with harmful statements within the generated responses. The adoption of different demographic personas can lead to unequal or biased treatment of users based on their demographic attributes \cite{sheng2021revealing}. With a similar outlook, Melchiorre et al. \cite{melchiorre2020personality} investigate how different user personalities affect the recommendations that they receive in a music recommender system.






\begin{table}
\begin{tabular}{p{2.7cm}|p{1.4cm}|p{1.45cm}}
\centering
\small
Bias Name & Rises From & First Affects \\ \Xhline{5\arrayrulewidth}
    Popularity      &      RS      &    DMS     \\ \hline
    User Log      &      CRS       &    Logs     \\ \hline
    Recomm. Evaluation & RS Training & RS\\ \hline
    Attribute     &      RS      &    DMS     \\ \hline
    Position &      DMS      &     NLG    \\ \hline
    Personalization &      RS      &     DMS    \\ \hline
    Context &      RS     &     User    \\ \Xhline{3.5\arrayrulewidth}
    Anchoring &     DMS       &     RS    \\ \hline
    Attribute Selection &      DMS      &     NLG    \\ \hline
    Human-AI Interaction &      User      &    NLU     \\ \hline
    Modality &      NLU      &     DMS    \\ \Xhline{3.5\arrayrulewidth}
    Defective Queries &      User      &     NLU    \\ \hline
    Cognitive &      NLG      &     User    \\\hline
    Unintended &      NLG (NLU)      &     User (DMS)    \\\hline
    Persona &      NLU (NLG)      &    DMS (DMS)     \\ \hline
\end{tabular}
  \caption{The various biases and their origins are listed, along with their effects on different modules of the CRS. Each of the biases can be traced back to a route in the architecture of a CRS  illustrated in Figure \ref{fig1:arch}. The start of route would be where the bias rises from and the initial effecting point of the bias would be the end of the route. This would enable us to navigate through the system in order to investigate the interplay of the different biases and their effects on each other.}
  \label{tab:table1}
\end{table}
\section{Limitations and Future Work}
\label{sec:future}

When conducting a survey on CRS biases, it's crucial to acknowledge some limitations that might impact the scope and depth of the findings. This paper acknowledges the potential relevance of natural language processing conferences as valuable sources of related references on biases \cite{wan2023biasasker}. However, these conferences were not extensively explored here. Additionally, focusing solely on papers published after 2019 might overlook earlier works providing historical context and a deeper understanding of CRS biases' evolution. While the survey aims for wide-ranging coverage, the references might not encompass all relevant literature on each bias. Nevertheless, the survey prioritizes breadth to establish a foundational understanding. Furthermore, lacking formal bias definitions and evaluation metrics may affect results, making their inclusion desirable for better consistency. Lastly, the study's limitations include the absence of experimentation on datasets, models, and presented mitigation methods. Despite these limitations, the survey offers valuable insights into CRS biases, paving the way for more in-depth research and robust approaches to address and mitigate biases in recommendation systems effectively.

In addition to addressing the above limitations, in future research endeavours, an exploration of the intersection between these biases presents a promising avenue. Delving into the intricate interplay of these biases of the system could offer a more comprehensive comprehension. Additionally, leveraging the insights from Table \ref{tab:table1}, a deeper understanding of how certain biases might synergize within specific system architectures could shed light on the propagation of bias and its effects on user interactions and decision-making processes.

\section{Conclusion}
\label{sec:conclusion}
This survey paper establishes a non-linear taxonomy of biases in conversational recommender systems. The investigation first covers biases in classic recommender systems and their adaptability to conversational recommender systems, highlighting their nature in conversational settings. Subsequently, biases in conversational recommender systems are explored in two parts: attribute-aware systems and topic-guided systems. By providing this taxonomy, the paper aims to aid researchers and developers in effectively addressing biases, ensuring fairness, transparency, and user trust in these systems' direct impact on user decision-making.

\section*{Acknowledgments}
This research was partially supported by the Canada CIFAR AI Chair program (Mila) and the Natural Sciences and Engineering Research Council of Canada (NSERC). 

\bibliography{sample-1col}

\end{document}